# Tailoring Physical Properties of Crystals through Synthetic Temperature Control: A Case Study for new Polymorphic NbFeTe$_2$ phases

Hanlin Wu[1], Sheng Li[1], Yan Lyu[2,3], Yucheng Guo[4,5], Wenhao Liu[1], Ji Seop Oh[4,6], Yichen Zhang[4], Sung-Kwan Mo[5], Clarina dela Cruz[7], Robert J. Birgeneau[4,6], Keith M. Taddei[7,*], Ming Yi[4,*], Li Yang[2,*] and Bing Lv[1,*]

1 Department of Physics, University of Texas at Dallas, Richardson, TX, 75080, USA

2 Department of Physics and Institute of Materials Science and Engineering, Washington University, St. Louis, MO, 63130, USA

3 Department of Physics, Nanchang University, Nanchang, 330031, China

4 Department of Physics and Astronomy, Rice University, Houston, TX, 77005, USA

5 Advanced Light Source, Lawrence Berkeley National Laboratory, Berkeley, CA, 94720, USA

6 Department of Physics, University of California, Berkeley, CA, 94720, USA

7 Neutron Scattering Division, Oak Ridge National Laboratory, Oak Ridge, TN, 37831, USA

To whom correspondence should be addressed: ktaddei@anl.gov, mingyi@rice.edu, yangli@wustl.edu, blv@utdallas.edu

**Abstract**

Growth parameters play a significant role in the crystal quality and physical properties of layered materials. Here we present a case study on a van der Waals magnetic NbFeTe$_2$ material. Two different types of polymorphic NbFeTe$_2$ phases, synthesized at different temperatures, display significantly different behaviors in crystal symmetry, electronic structure, electrical transport, and magnetism. While the phase synthesized at low temperature showing behavior consistent with previous reports, the new phase synthesized at high temperature, has completely different physical properties, such as metallic resistivity, long-range ferromagnetic order, anomalous Hall effect, negative magnetoresistance, and distinct electronic structures. Neutron diffraction reveals out-of-plane ferromagnetism below 70K, consistent with the electrical transport and magnetic susceptibility studies. Our work suggests that simply tuning synthetic parameters in a controlled manner could be an effective route to alter the physical properties of existing materials potentially unlocking new states of matter, or even discovering new materials.

# I. INTRODUCTION

The unique atomically thin 2D van der Waals (*vdW*) structures offer a remarkable platform for investigating the interplay between the spin, charge, orbital, and lattice degrees of freedom. [1–6] They also give rise to new physical phenomena including novel intrinsic magnetism and frustrated magnetism in the 2D atomic limit. The 2D magnetism was first discovered in $CrI_3$ [7] and $Cr_2Ge_2Te_6$ [8] despite predictions by the Mermin-Wagner theorem that prohibit long-range magnetic order at finite temperatures in isotropic 2D systems. Furthermore, 2D magnetism has also been achieved in the layered transition metal dichalcogenides (TMDs) such as $CrTe_2$ [9] and $VSe_2$ [10]. The combination of electronic structure and magnetism make the TMDs more interesting potentially hosting novel quantum phenomena. [11] The coexistence of multiple stable phases for TMDs with slight differences in the interatomic distance and coordination environment causing significant changes in their physical properties, has been rather appealing, especially for the metastable phases such as the $T_d$, 1T, 1T' and 1T''' phases. [12–14] Therefore, exploring new 2D *vdW* magnetic materials which are structurally and chemically akin to TMDs, or magnetically intercalating metastable TMD phases, will be a fertile field that could open new research avenues towards emergent phenomena.

In order to discover new phases for intercalated TMDs and new 2D *vdW* magnets, modulating the synthetic parameters such as growth temperatures and fluxes, has been found to be very effective to tune the physical properties and even lead to the discovery of new materials [15–19]. For instance, more than ten unique structural types have been discovered in the ternary copper chalcogenide system by systematically varying the temperature and flux ratios without altering the proportions of starting materials. [20–22] By simply changing the flux and synthetic temperature, new polymorphic $BaCu_2As_2$ phase with intergrowth feature and new $BaCu_6Sn_2As_{4-x}$ phases are identified in the copper pnictide system. [17,23] Similarly, significant changes of physical properties have been reported in the layered $ZrTe_3$ crystals synthesized at different temperatures. [18] The low-temperature-synthesized crystals display a charge density wave (CDW) at 70K while high-temperature-synthesized crystals, with atomic disorders at Zr and Te sites, show suppression of CDW order and bulk superconductors at 4K.

NbTe$_2$ is a non-magnetic 1T' TMD phase, which exhibits the coexistence of CDW order with a transition temperature above 550K and superconductivity below 0.75K. [24–26] In the course of our intercalation studies, where we introduced Fe, Co and Ni ions aiming to induce magnetic orders in this system, we found two distinct NbFeTe$_2$ phases at different synthetic temperatures, which exhibited drastic changes in crystal symmetry and physical properties. The low-temperature synthesized phase (LT phase) possessed an orthorhombic structure which has been reported previously. [27,28] It can be treated as the interstitial sites filled T$_d$ phase, and the experimental results suggested it is an Anderson insulator with spin glass behavior, consistent with previous studies. [28] The high-temperature synthesized phase (HT phase) crystalizes in a monoclinic crystal structure and displays a clear ferromagnetic order with transition temperature T$_c$ around 70K. Consistent with the ferromagnetic order, a large negative magnetoresistance and anomalous Hall effect are observed in this system. Furthermore, we found that the LT phase can be transformed into the HT phase through simple thermal annealing. Our results demonstrate an effective yet simple approach for examining the effects of synthetic parameters in a controlled manner, which can not only lead to the discovery of new quantum materials, but also provide new insights into their magnetic, transport properties and functionalities of existing materials.

## II. EXPERIMENTAL SECTION

Single crystals of both HT and LT NbFeTe$_2$ were synthesized using the chemical vapor transport method using I$_2$ as transport agent with the starting materials Nb powder (99.9%, Alfa Asear), Fe powder (99.99%, Alfa Asear), and Te lumps (99.999+%, Alfa Asear). All the synthesis procedures were carried out within a purified Ar-atmosphere glovebox with total O$_2$ and H$_2$O levels <0.1 ppm. Chemical stoichiometric elements of Nb, Fe and Te were loaded into the quartz tube with a small amount of I$_2$ (1 mg/cm$^3$). The quartz tubes were then put into the tube furnace with different temperature profiles. The HT phase with typical size of 3×3×0.2 mm$^3$ were obtained at the source side for two weeks reaction with temperature profile of 1000 °C (source)/900 °C (sink), while the LT phase with typical size of 1×1×0.1 mm$^3$ were obtained for the same two weeks reaction with temperature profile of 750 °C(source)/650 °C (sink). Both crystals are plate-like with shinning metallic luster surfaces. Different synthetic parameters have also been tested to figure out the growth windows and optimize the growth conditions for the HT phase. The HT phase could also be synthesized with temperature gradient of 950 °C (source)/850 °C (sink) or 900 °C (source)/800 °C (sink) with prolonged growth time over a month yet with smaller size crystals. In these growth conditions, we did not observe coexisting HT and LT phases. The crystals of LT phase could be converted to HT phase by post-annealing the crystals at 950 °C for a week.

Powder X-ray diffraction (XRD) was performed using a Rigaku Smartlab diffractometer with Cu K$_\alpha$ radiation. Single-crystal X-ray analysis was performed using a Siemens SMART diffractometer equipped with a CCD area detector and monochromatic Mo K$\alpha$1 radiation ($\lambda$ = 0.71073 Å). The collected dataset was integrated with Bruker SAINT and scaled with Bruker SADABS (multi-scan absorption correction). [29] A starting model was obtained using the direct method in SHELXT [30] and atomic sites were refined anisotropically using SHELXL2014. The composition of all crystals was confirmed by SEM energy-dispersive X-ray spectroscopy (SEM-EDX) using Zeiss EVO LS 15 SEM with accelerating voltage of 20 keV. The data was collected on several crystals with at least five measured points for each crystal to confirm the homogeneity and accurate composition of the crystals. The electrical resistivity, Hall effect and specific heat data was performed in the Quantum Design Physical Property Measurement System (PPMS).

Temperature and field dependent magnetization data was measured on the Quantum Design DynaCool system.

Calculations are performed by using first-principles density functional theory (DFT) with the Perdew–Burke-Ernzerhof (PBE) functional and a kinetic energy cutoff of 400 eV, as implemented in *Vienna Ab initio Simulation Package* (VASP) [31,32]. The first Brillouin zones are sampled with $6 \times 10 \times 8$ k-point meshes. vdW interactions are adopted by the DFT-D3 method [33]. On site Hubbard interaction is adopted by the Dudarev scheme [34], with a range of U from 0 to 3 eV for Fe ions. For geometry optimizations, all atoms are fully relaxed until the residual force per atom is less than 0.01 eV/Å. Spin-orbital coupling (SOC) is included in the calculations.

Neutron powder diffraction (NPD) measurements were performed on the HB-2A powder diffractometer of Oak Ridge National Laboratory's (ORNL) High Flux Isotope reactor (HFIR) [35]. Patterns were collected between 1.5 and 125 K; and under an applied magnetic field between 0 and 4 T using a cryomagnet. An incident wavelength of 2.41 Å and pre-monochromator, pre-sample and pre-detector collimator settings of open, 21' and 12' respectively. Full patterns were collected with 4 hr count times and order parameter-like scans were collected on magnetic Bragg peaks by moving the detector to the relevant 2θ and collecting on warming. Single crystal neutron diffraction data were collected on the WAND$^2$ diffractometer of ORNL's HFIR. Data were collected in the H0L scattering plane between 1.5 and 200 K using a cryostat with a vertically focused incident beam of 1.48 Å. Rietveld refinements were performed using the Fullprof software suite [36]. Representational analysis was performed using SARAh. [37]

ARPES measurements were performed using a Scienta R4000 electron analyzer at Beamline 10.0.1 of the Advanced Light Source (ALS) with an energy resolution and angular resolution of 12 meV and 0.3°, respectively. The samples were cleaved *in-situ* and measured with the base pressure below $4\times10^{-11}$ Torr at 15K.

# III. RESULTS AND DISCUSSION

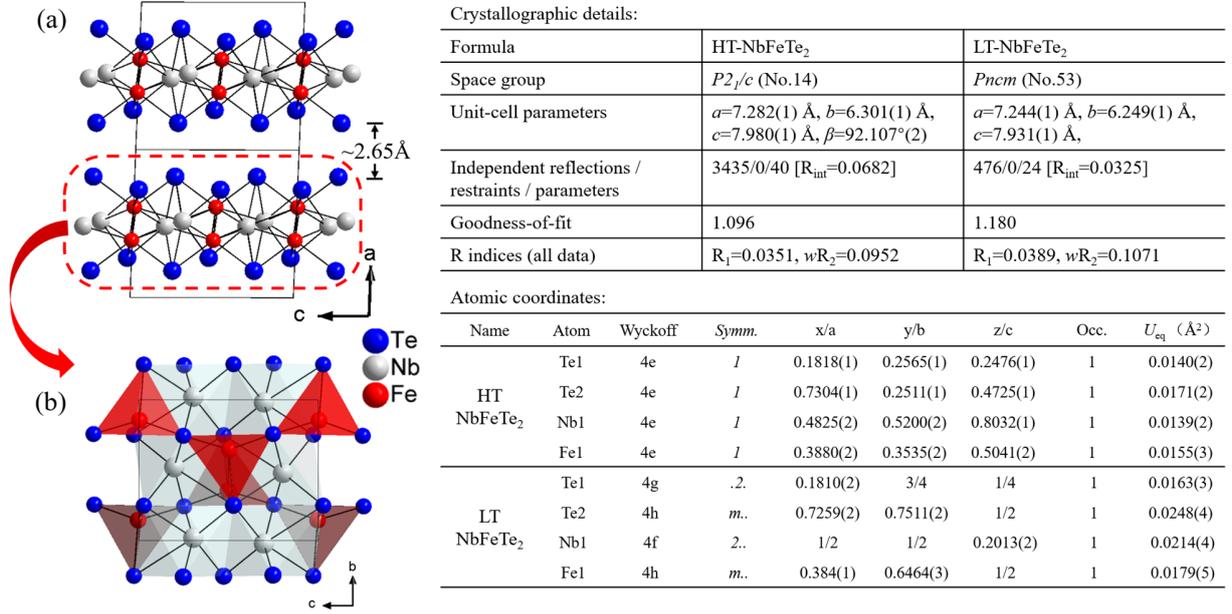

FIG. 1: (a) Crystal structure of HT phase of NbFeTe$_2$ projected along the *b* direction showing the layered structure; (b) The in-plane NbFeTe$_2$ layer projected along *a* direction highlighting NbTe$_6$ octahedra and FeTe$_4$ tetrahedra in the *bc* plane. (c) Crystal refinement details and atoms coordination for both HT phase and LT phase of NbFeTe$_2$.

Both HT and LT crystals show only the NbFeTe$_2$ phase and no other impurity elements present from chemical analysis in SEM-EDX (Fig. S1). However, significant differences in crystal symmetry were This will be subject to our investigation in the future. observed in X-ray single crystal diffraction. As shown in Fig. 1, both HT and LT phases have layered structures with interlayer distances of ~2.65 Å. The layers [Fig. 1(b)] consist of NbTe$_6$ octahedra, which are face sharing along the *b* axis and edge sharing along *c* axis, with Fe atoms at the interstitial sites forming FeTe$_4$ tetrahedra. Each Fe atom form a dumbbell-like motif with another Fe atoms, and each dumbbell is connected through extra Te atoms. The LT phase is found in the orthorhombic system with space group *Pncm* (No. 53). The refined unit cell parameters are *a* = 7.244(1) Å, *b* = 6.249(1) Å, and *c* = 7.931(1) Å, which are consistent with previously reported phase [27]. The HT phase crystallizes in the monoclinic system with space group *P2$_1$/c* (No. 14), and the refined unit cell parameters are *a* = 7.282(1) Å, *b* = 6.301(1) Å, and *c* = 7.980(1) Å and β = 92.11°. We note that we intentionally use the non-standard space group *Pncm* rather than the standard #53 *Pmna* for the refinement so that one can directly compare the difference between the HT and LT phases. The detailed refinement results are shown in Fig. 1(c), with the final CIF files for both compounds provided in the supplemental information. Both phases have two distinct crystallographic sites for Te, one distinct site for Fe and Nb atoms, respectively. Most of the Nb-Te and Fe-Te distances are similar with each other between HT phase and LT phase [2.764(1)- 2.863(1) Å for Nb-Te and 2.551(2)- 2.646(2)Å for Fe-Te], and are similar to bond distances in the $Nb_2SiTe_4$ (2.845-2.965 Å) [38], $NbTe_2$ (2.695-2.885Å) [39], $Nb_3Sb_2Te_5$ (2.894-2.927 Å) [40] and $FeTe_2$ (2.552- 2.564) Å [41]. The major difference between the two phases lies in the placement of the Nb atom, where it changes from the higher symmetric *4f* site [½, ½, 0.2013(2)] for the LT phase, to a lower *4e* site [0.4825(4), 0.5200(4), 0.1968(4)]. As such, it causes distortion on the $NbTe_6$ octahedra where the two Nb-Te2 distances are changed from 3.278(1) Å in LT phase to 3.012(1) Å and 3.667(2) Å in the HT phase with associated severe Te-Nb-Te angle changes. The 3.667 Å Nb-Te2 distance suggest non-bonding between the two atoms, and thus cause the more severe distortion on the $NbTe_6$ octahedra packing, resulting in the broken structural symmetry and its transformation from orthorhombic in LT phase to monoclinic HT phase.

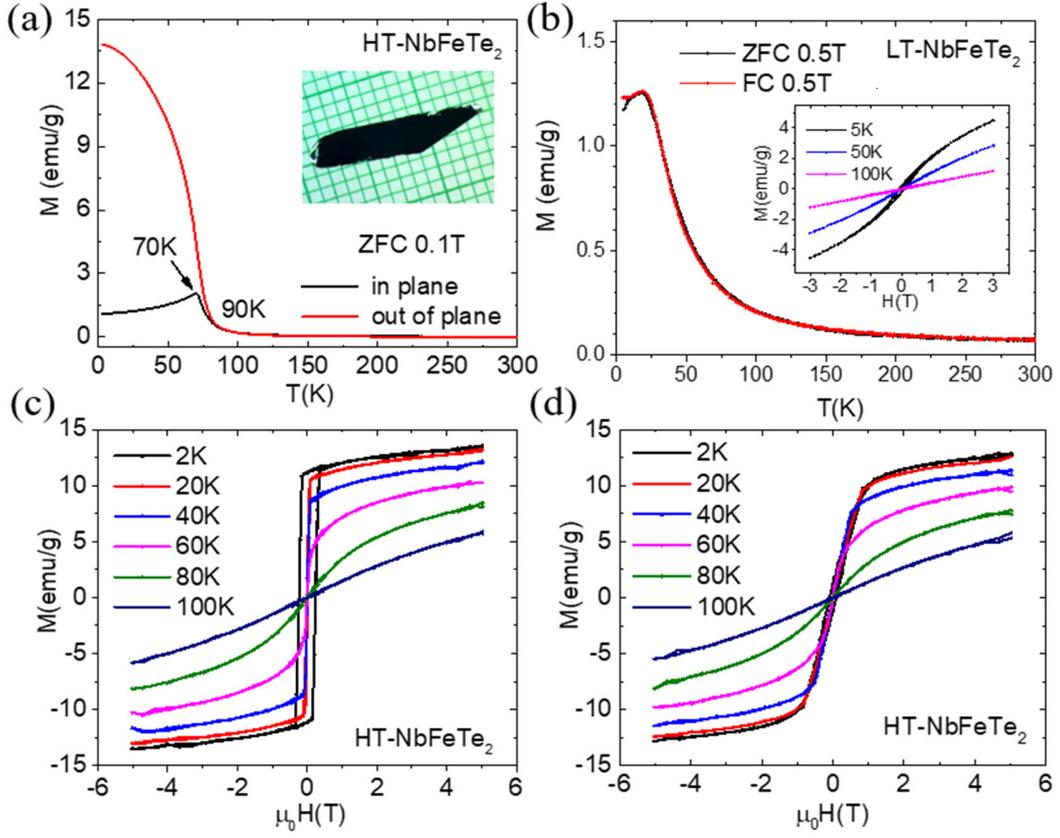

**FIG. 2:** (a) Temperature dependent of magnetization with magnetic field along and perpendicular to crystal plane under magnetic field 0.1T for HT NbFeTe$_2$. Inset is the image of single crystal of HT NbFeTe$_2$ on a millimeter-scale sheet. (b) Temperature dependent of magnetization of LT NbFeTe$_2$. Inset is magnetic hysteresis loops at different temperature. Isothermal magnetic hysteresis loops of HT NbFeTe$_2$ with field direction (c) perpendicular and (d) parallel to the crystal layers.

To explore the influence of the structural difference between HT phase and LT phase, we investigate the magnetic and transport properties, and surprisingly find that these two phases show completely different behaviors. Temperature dependent magnetization of HT phase is shown in Fig. 2(a). By applying a magnetic field along different orientation, we observe a distinct magnetization behavior for HT phase, where a typical FM behavior with magnetic field perpendicular to the layer and a cusp at 70K with magnetic field parallel to the layer. Such behavior is also observed in other layered magnets which indicate the magnetic moment is perpendicular to the layer. [42,43] The magnetization curves overlap above 90K, and the splitting of magnetization between $T_c$ and 90K suggests strong spin fluctuations within this temperature range. The LT phase, clearly show no magnetic order at high temperature with a spin glass magnetic transition occurred at low temperature (~15 K) [Fig. 2(b)]. This is consistent with the previous report [28].

By applying a magnetic field perpendicular and parallel to the layer direction of HT phase, magnetic anisotropic behaviors are observed as shown in Fig. 2(c) and 2(d). Based on the magnetization hysteresis (MH) loops at different temperatures, the *a*-axis is recognized as the easy axis for magnetization, because the saturation field along H // *a* ($H^a$ ~ 3.5 kOe) is far below that of H // *bc* ($H^{bc}$ ~ 13 kOe). This anisotropy is further highlighted by the difference in saturated magnetic moments calculated from Fig. 2(c) and 2(d), which are estimated to be $\mu_S^a=0.95\mu_B$ and $\mu_S^{bc}=0.89\mu_B$. Additionally, the nonlinear magnetization loops observed at 80K in both directions are consistent with temperature dependent magnetization in Fig. 2(a), which are likely due to strong spin fluctuations.

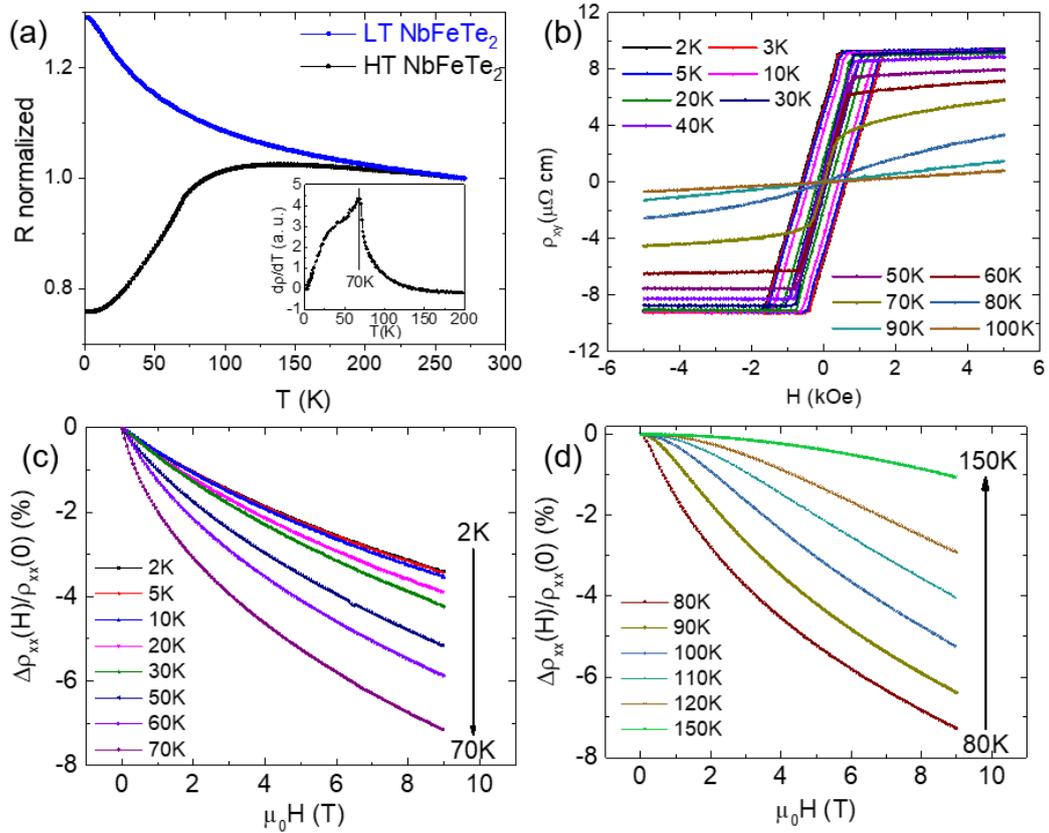

**FIG. 3:** (a)Temperature dependent normalized resistivity data of LT and HT NbFeTe$_2$. Insert is first derivative of HT resistivity curve. (b) Hall resistivity of HT NbFeTe$_2$ at different temperature. Magnetoresistance of HT NbFeTe$_2$ at (c) 2-70 K and (d) 80-150 K.

Besides the LT and HT phases exhibiting different magnetic ground states, the electrical properties also show distinct behaviors between the two phases. Temperature dependent resistivity data for the HT and LT phases are shown in Fig. 3(a). Consistent with previous results, the resistivity of the LT phase shows a semiconducting trend at low temperature due to the strong localization, resulting in the LT phase likely being an Anderson insulator. The Anderson localization may arise from the small Fe vacancies [44]. Resistivity of the HT phase increases slightly as temperature decreases at first, and then decreases with further decreasing of the temperature, showing a metallic ground state. A broad peak at 70K in the first derivative of the resistivity data [inset of Fig. 3(a)] is observed, consistent with the magnetic transition of HT phase at 70K. To exclude that the FM transition and resistivity anomaly in the HT phase originates from a structural transition, we performed both temperature dependent single crystal diffraction down to 80K and Neutron diffraction at low temperature (discussed later), and no such transition is observed. This anomaly might be associated with a Lifshitz transition, similar to that of ZrTe$_5$ [45,46]. We indeed observe an anomalous Hall effect (AHE) up to 80 K in Fig. 3(b) for the HT phase, which slightly exceeds the FM transition temperature at 75K but is consistent with our suggestions of strong spin fluctuations persisting up to ~90 K from magnetic anisotropic measurements. We then investigate magnetoresistance (MR) for the HT phase at different temperatures. A large negative MR (nMR) is observed from 2K to 150K as shown in Fig. 3(c) and 3(d). The nMR curvature follows parabolic behaviors, where the MR value at 9T first increases and then decreases as the temperature increases. A crossover of nMR at ~ 70K is observed, which corresponds to the FM transition and could be attributed to the electron scattering by strong spin fluctuations near the magnetic transition [47].

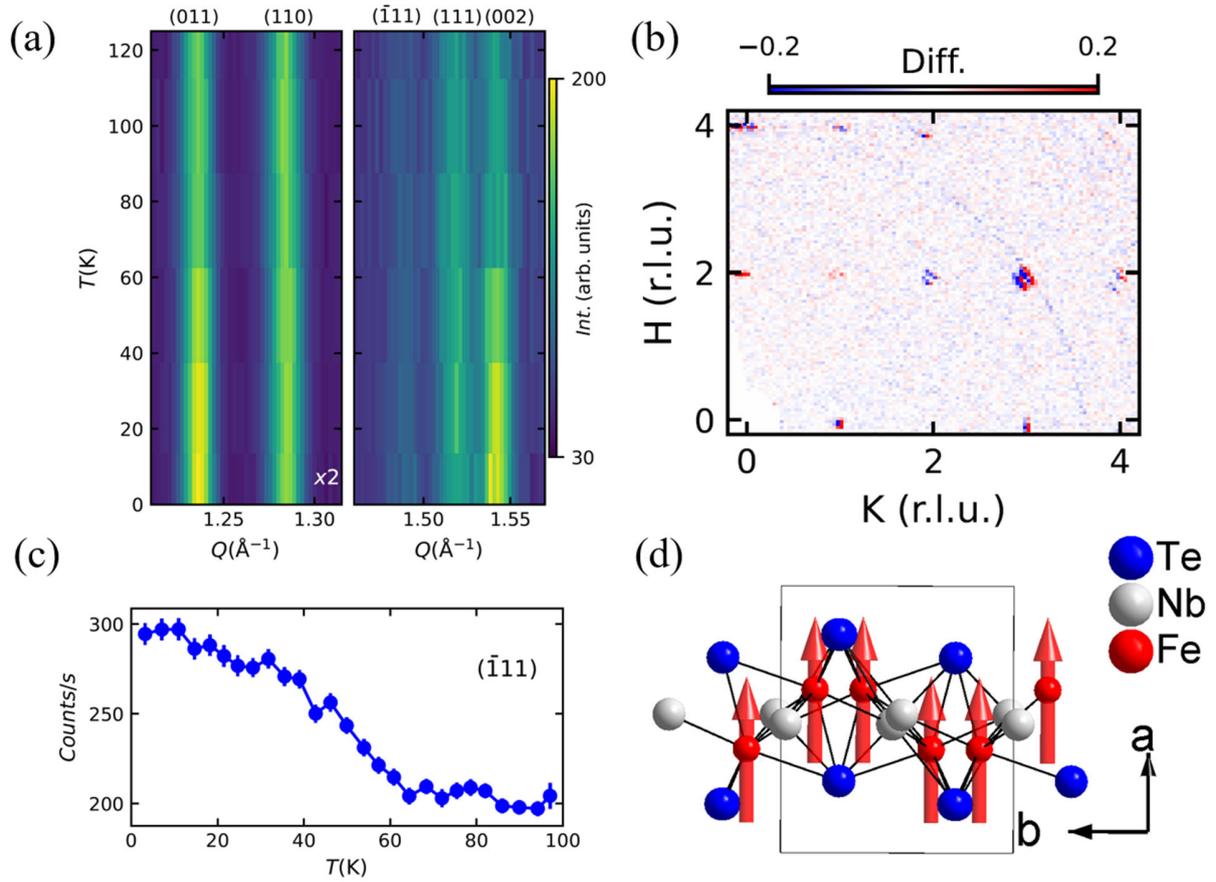

**FIG. 4:** (a) Neutron powder diffractogram of HT NbFeTe$_2$. (b) Order parameter scan collected while warming the sample. (c) Difference plot of low (10 K) and high (100 K) neutron scattering intensities of a single crystal sample. (d) Magnetic structure determined from representational analysis using neutron diffraction data.

To unambiguously elucidate the magnetic order in the HT NbFeTe$_2$ phase, we perform NPD measurements. NPD patterns and best fit models from Rietveld refinements for data for 125K and 2K are shown in supplemental Fig. S2. In Fig. 4(a), temperature dependent neutron powder diffractogram shows the scattering intensity for a series of low Q peaks at low temperatures. Here a change in scattering intensity is clearly seen which coincides with the signal at 70K, observed in both the magnetization and resistivity measurements. As no such intensity change is observed in the XRD we attribute the scattering to a magnetic origin. To better characterize the transition, we collect the intensity of the $\bar{1}11$ peak as a function of temperature upon warming [Fig. 4(b)]. The peak intensity increases with decreasing temperature when temperature is below 70K. The $\bar{1}11$ peak is chosen due to its seemingly minimal contribution from nuclear scattering as seen in its nearly becoming background equivalent about 70 K in Fig. 4(a). To carefully check for any weak additional magnetic scattering and help discriminate between potential magnetic symmetries, single crystal neutron diffraction is collected in the (H 0 L) plane at 10K and 100 K and then plotted as a difference map in Fig. 4(c). As seen, there is some difference in the intensities seen at integer positions, which is consistent with FM ordering, but no additional scattering is observed. With this information, magnetic structure solution is performed using representational analysis to consider all potential magnetic structures allowed by a ferromagnetic k=000 ordering vector, the Wyckoff position of the Fe site and the crystallographic space group, as shown in Table S1. Of the potential magnetic structures, the best fit model was found to have purely ferromagnetic components along *a* axis can be characterized via representational analysis as the $\Gamma_3$ irreducible representation of the nuclear space group and the (0,0,0) ordering vector, which corresponds to the magnetic space group $P2_1'/c'$, as shown in Fig. S2. In Fig. 4(d), the magnetic structure is shown with all the Fe magnetic moments along *a* axis as expected from the temperature dependent magnetization measurement in Fig. 2a. The refined magnetic moment of Fe is 0.4 $\mu_B$, which is smaller than the value estimated from magnetic measurements and can be consistent with itinerant ferromagnetism as seen in systems such as $Fe_3GeTe_2$. [48] We note that although our solution only contains aa moment component along *a*, this is not an explicit constraint of the $\Gamma_3$ irrep. As shown in table S1, $\Gamma_3$ has three basis vectors, one each for the three crystallographic directions. In our analysis, the best fit was produced with a model where only the basis vector describing a FM moment along *a* was allowed to have a non-zero contribution. None the less, as the moment is allowed by symmetry to have non-zero components along *b* and *c* we cannot completely rule them out. However, we can put an upper limit on their value based on the sensitivity of our NPD measurements at < 0.1μB. Additional measurements were performed under an applied field under both field cooled (FC) and zero-field cooled (ZFC) procedures. No additional peaks nor meaningful change in the peak intensities was observed indicating the absence of a metamagnetic transition up to 4 T.

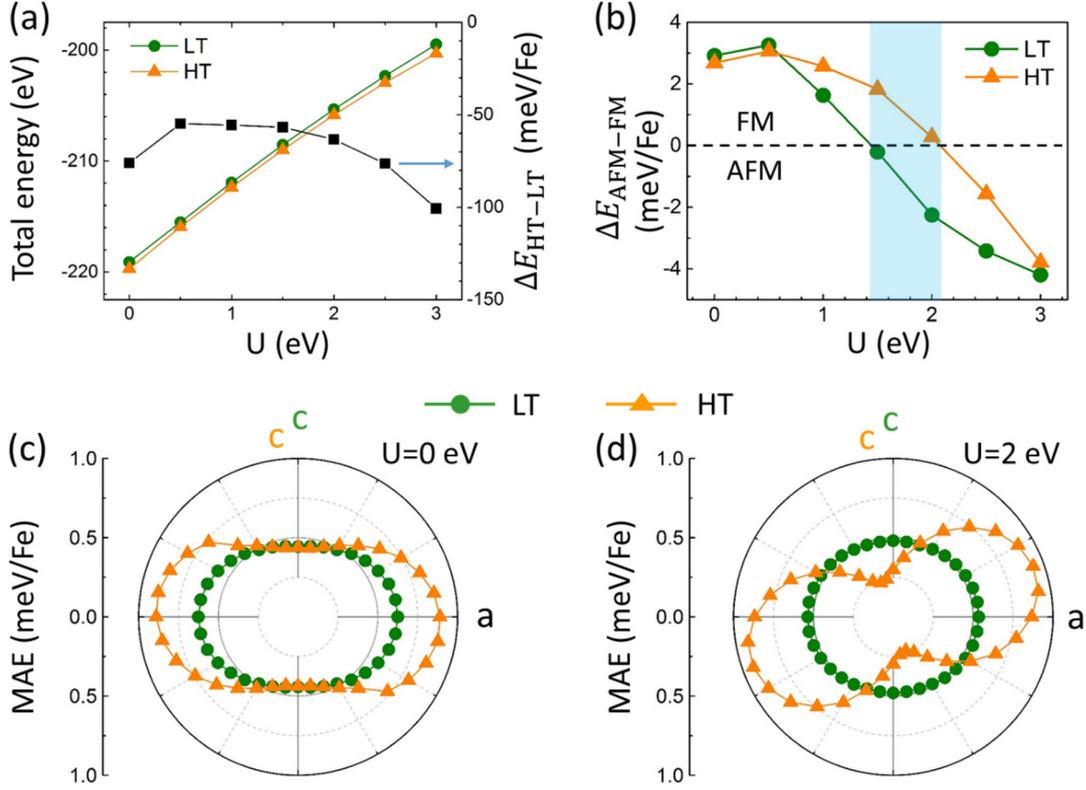

**FIG. 5:** (a) DFT-calculated total energy and energy difference between LT and HT NbFeTe$_2$. (b) Interlayer coupling as a function of U of Fe ions. Magnetic anisotropic energy in the *ac* plane for LT and HT NbFeTe$_2$, calculated by (c) U=0 eV and (d) U=2 eV.

To fully understand the experimental results, we investigate the electronic and magnetic states of HT and LT phases of NbFeTe$_2$ by using DFT calculations. The energy bands and projected density of states calculated by DFT+U with different U value are shown in Fig. S3 and S4. As shown in Fig. 5(a), the HT phase has lower total energies than the LT phase, which is about 50-100 meV/Fe lower in the range of 0-3 eV of the value of Hubbard interaction (U) of Fe ions. This supports the observation that LT phase can transfer to HT phase by thermal annealing process. The local magnetic moments are found mostly from the Fe ions. For example, the magnetic moments of Fe and Nb ions are 2.55 and -0.32 $\mu_B$, respectively, calculated by U = 2.0 eV for Fe ions. Calculations of possible magnetic states suggest that the ground state of intralayer magnetic coupling is FM between Fe ions and AFM between Fe and Nb ions for both LT and HT phases. These intralayer coupling states are strong and robust with respect to different values of U. On the other hand, the ground state of interlayer coupling is sensitive to the value of U of Fe ions, as shown in Fig. 5(b), where the critical values of U for interlayer FM/AFM transition are different for LT and HT phases. In the range of 1.4-2.1 eV of U, the LT phase is interlayer AFM, while the HT phase is interlayer FM, which agrees with the measured FM state of the HT phase.

To explain the different magnetic orders of HT and LT phases, the magnetic anisotropic energy (MAE) with different values of U is calculated and plotted in Figs. 5(c) and 5(d). The magnetic easy axis is nearly along the out-of-plane direction (the *a* axis) for both HT and LT phases. However, when comparing in-plane directions (along *c* axis) with U=0 eV, the HT phase is more anisotropic than the LT phase [Fig. 5(c)]. And as increasing value of U to 2 eV, the LT phase becomes more isotropic, while the HT phase becomes more anisotropic in the *ac* plane [Fig. 5(d)]. As it is believed that MAE is necessary to induce the long-range magnetic orders in layered magnetic materials at finite temperature [8], the significant MAE may contribute to the observed FM order in the HT phase while the nearly isotropic MAE in the LT phase may result in the spin glass state, instead of the long-range order.

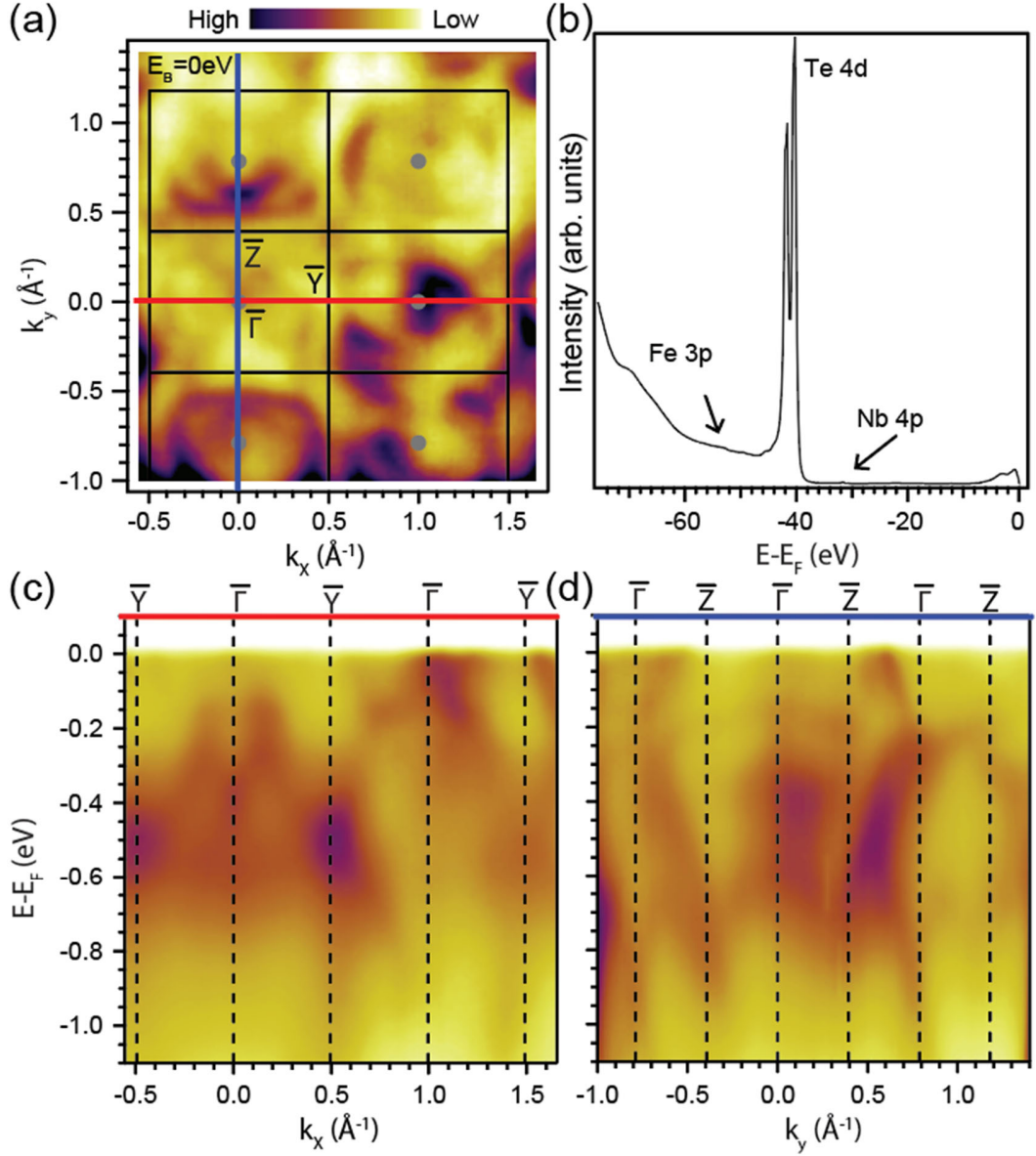

**FIG. 6:** (a) Fermi Surface of HT NbFeTe$_2$. The Brillouin zones are labeled in black. (b) Experimental angle-integrated photoemission spectra. (c) Band dispersions along $\bar{Y}$-$\bar{\Gamma}$-$\bar{Y}$, denoted by the red line in (a). (d) Band dispersions along $\bar{Z}$-$\bar{\Gamma}$-$\bar{Z}$, denoted by the blue line in (a). All measurements were performed with 78 eV photons at 15K.

To investigate the electronic structure, we have carried out ARPES measurements of HT NbFeTe$_2$ in the FM state (Fig. 6). Fermi surface map in Fig. 6(a) demonstrates pronounced matrix element effects, which are further revealed in the band dispersions along the high symmetry directions in Fig. 6(c)-(d). In the cuts, the band dispersions along $\bar{Z}$- $\bar{\Gamma}$ (or $\bar{Y}$- $\bar{\Gamma}$) are normally expected to be identical with the band dispersion along the same cut in different Brillouin zones due to the translational symmetry of the lattice. Here, however, the bands exhibit significant intensity variations due to the strong photoemission matrix element effect. Its origin needs further investigation. The core levels in the angle-integrated photoemission spectra in [Fig. 6(b)] confirms the existence of Tellurium, Niobium, and Iron in the compound. The dispersive bands at the Fermi level suggests the system is metallic in the FM state. This observation is consistent with the electrical transport results and stands in stark contrast to the LT NbFeTe$_2$ where flat bands emerge at the Fermi level due to Anderson localization and result into an insulating behavior in the electrical transport measurements [28].

**IV. CONCLUSION**

In conclusion, we have presented a case study on discovery of a high-temperature polymorphic phase of layered $NbFeTe_2$ by simply modifying the synthetic temperatures. Compared to previously reported $NbFeTe_2$ with spin glass transition, this new polymorphic HT $NbFeTe_2$ phase has lower crystal symmetry and shows completely different physical properties. Electrical transport, magnetic susceptibility and Neutron diffraction studies show a clear long range out-of-plane ferromagnetic transition at 70K. ARPES study confirm a metallic electronic structure in the FM state. HT $NbFeTe_2$ also displays a metallic behavior with negative MR over the whole temperature range, and AHE occurs below ferromagnetic transition temperature. The electronic and magnetic states of two different phases of $NbFeTe_2$ have been investigated while the simulation results agree with the measurements. The first-principles calculation suggests that the variation of MAE could be the origin for observed different magnetic orderings in LT and HT phases. It appears that synthetic parameters by both ex-situ temperatures/post-annealing or in-situ synthesis/diffraction combination in a controlled manner could be fruitful directions to explore for materials discovery in the future.

**Acknowledgements:**

This work at the University of Texas at Dallas is supported by the US Air Force Office of Scientific Research (AFOSR) Grant no. FA9550-19-1-0037, National Science Foundation (NSF) (DMREF-1921581) and Office of Naval Research (ONR) grant no. N00014-23-1-2020. Part of our measurement facilities acknowledge the support from the AFOSR Defense University Research Instrumentation Program (DURIP) grant no. FA9550-21-1-0297. The ARPES work at Rice University was supported by the Robert A. Welch Foundation Grant No. C-2175 and the Gordon and Betty Moore Foundation's EPiQS Initiative through grant No. GBMF9470. This research used resources of the Advanced Light Source, which is a DOE Office of Science User Facility under contract no. DE-AC02-05(c)H11231. Yucheng Guo was supported in part by an ALS Doctoral Fellowship in Residence. L.Y. is supported by NSF DMREF DMR-2118779. The simulation used Anvil at Purdue University through allocation DMR100005 from the Advanced Cyberinfrastructure Coordination Ecosystem: Services & Support (ACCESS) program, which is supported by National Science Foundation grants #2138259, #2138286, #2138307, #2137603, and #2138296. A portion of this research used resources at the High Flux Isotope Reactor, a DOE Office of Science User Facility operated by the Oak Ridge National Laboratory.

Author Information

Keith M. Taddei: ORCID 0000-0002-1468-0823

Sung-Kwan Mo: ORCID 0000-0003-0711-8514

Yichen Zhang: ORCID 0000-0001-6792-330x